\documentstyle[12pt]{article}
\begin{document}
\begin{center}
{\bf BOSONIZATION AND DUALITY IN 2+1-DIMENSIONS:\\
 APPLICATIONS IN
GAUGED MASSIVE THIRRING MODEL}\\
\vskip 2cm
Subir Ghosh
\footnote {Email address: <subir@boson.bose.res.in>}\\
Physics Department,\\
Dinabandhu Andrews College, Calcutta 700084,\\
India.\\
\end{center}
\vskip 2cm
{\bf Abstract:}\\
\vskip .5cm
Bosonization of the gauged, massive Thirring model in 2+1-dimensions
produces a Maxwell-Chern-Simons gauge theory, coupled to a dynamical,
massive vector field. Exploiting the Master Lagrangian formalism,
two dual theories are constructed, one of them being a gauge theory. The
full two-point functions of both the interacting fields are computed in
the path integral quantization scheme. Furthermore, some new dual models,
derived from the original master lagrangian and valid for different
regimes of coupling parameters, are constructed and analysed in details.

\newpage
{\bf Introduction:}\\
\vskip .5cm
The notion of duality has played a key role in the common goal of
unification in physics. Broadly, two particular models are said to be
dual to each other if in some sense they are equivalent to each other
as regards to, {\it e.g.} effective action, Green's function, etc.. One
way of obtaining a pair of dual systems is to consider the so called
Master lagrangian \cite {djt} of a model of larger phase space, from
which each of the dual models are derivable. The presence of the Master
lagrangian assures the equivalence of the said (dual) models. The other
more direct way is to derive one model from another one - such as
Bosonization (in 1+1-dimensions), where a fermionic model, {\it e.g.} the
Massive Thirring Model (MTM) was shown to be dual to the (bosonic) sine-Gordon
model \cite {cm}.

Recently there has been a marriage of sorts between these two approaches
in higher (2+1) dimensions. Early works \cite{lm} in 2+1-dimensional
bosonization were not completely satisfactory, as the fermion-boson
mapping was plagued with non-locality. However, the programme received
an impetus from later works \cite{fsb}, where a local bosonic action
for the MTM is obtained at the cost of truncating the Seelay expansion
of the fermion determinant at one loop. The small parameter in the expansion
is the inverse of fermion mass.

In a recent paper \cite{sg}, we have been able to bosonize the Gauged
Massive Thirring Model (GMTM).
The spectrum of the bosonic model, a gauge theory, was obtained in
a classical setup \cite{sg}. The present work is devoted to a quantum
analysis of the
bosonized GMTM. Exact two-point functions of the GMTM are derived.
The concept of duality is utilised in two ways: On the one hand to generate
the full propagators of GMTM and on the other hand, to establish the
equivalence between several (manifestly) gauge invariant models with
non-invariant ones. The latter phenomenon is reminiscent of the duality
between abelian self-dual and Maxwell-Chern-Simons models \cite{dr}, and
non-abelian self-dual and Yang-Mills-Chern-Simons models \cite{bbg}. The
rest of this section is devoted to a brief elaboration of the computational
scheme adopted.

In section {\bf II}, the GMTM and its bosonized version is stated from
previous works \cite{sg}. The lagrangian of GMTM, $L_F(\psi, A_\mu,m,e,g)$
consists of the following parameters: the fermion ($\psi $) mass $m$, the gauge
field ($A_\mu$) coupling $e$ and the Thirring coupling $g$. The bosonized
version has $L_B(B_\mu,A_\mu,m,e,g)$ where $B_\mu$ is an auxiliary vector
field introduced to linearize the Thirring interaction. The latter model is
our Master Lagrangian (ML). Note that the duality between $L_F$ and $L_B$
is not exact, for reasons mentioned earlier.

In section {\bf III} we derive the dual partition functions $Z(A_\mu)$
and $Z(B_\mu)$, via integrating out $B_\mu$ and $A_\mu$ (in the Lorentz
gauge)
from ML respectively. This produces an exact duality between $Z(A_\mu)$, a
gauge invariant theory, and $Z(B_\mu)$, a non-invariant one.

The dual system helps us to compute the full propagators $<A_\mu A_\nu>$
and $<B_\mu B_\nu>$ of ML. Limiting values of the parameters are studied
to show the consistency of the above framework. This is the content of
section {\bf IV}.

Section {\bf V} comprises of some more dual models, one of the pairs
(of a dual system)
being a gauge theory. All these models are derived from the starting ML,
the distinction being the relative strengths of the couplings $e$ and $g$.
Once again Green's functions of these models are derived.

We end the paper with a brief conclusion in section {\bf VI}.
\vskip .5cm
{\bf II. Bosonized model}\\
\vskip .5cm
Our starting point is the gauged massive Thirring model,
\begin{equation}
L_F=\bar\psi i\gamma^\mu (\partial_\mu-ieA_\mu)\psi -m\bar\psi\psi
+{g\over 2}\bar\psi\gamma^\mu\psi\bar\psi\gamma_\mu\psi.
\label{eqlf}
\end{equation}
Here $A_\mu$ is an external, abelian gauge field and $g$ is the Thirring
coupling constant. The system has abelian gauge invariance under,
$$\psi\rightarrow exp(ie\alpha),~~~A_\mu\rightarrow A_\mu
+\partial_\mu\alpha.$$
Notice that no kinetic terms for $A_\mu$, ({\it e.g.} Maxwell or Chern-Simons term),
are present in (\ref{eqlf}) to give dynamics to the photon. However this is
no restriction since we will study the bosonized version of ({\ref{eqlf}),
where the kinetic terms corresponding to $A_\mu$ will be automatically
generated in the Seelay expansion of the fermion determinant. Hence the
coefficients of the starting kinetic terms, (if present), will get
renormalized anyway after bosonization.

We now introduce an auxiliary field $B_\mu$ to linearize the Thirring
interaction,
\begin{equation}
L_F=\bar\psi i\gamma^\mu [\partial_\mu-i(B_\mu+eA_\mu)]\psi -m\bar\psi\psi
-{1\over{2g}}B^\mu B_\mu.
\label{eqlfb}
\end{equation}
The partition function to be evaluated is
\begin{equation}
Z_B=\int {\cal D}(\psi\bar\psi)exp(i\int L_F).
\label{eqbos}
\end{equation}
Unlike in 1+1-dimensions, the fermion determinant is non-local and
conventionally one computes the one loop expression in a power series with
inverse fermion mass, ({\it i.e.}${1\over m}$), as the small parameter. Keeping
the $m$-independent and $O({1\over m})$ terms only we obtain,
\begin{equation}
L_B=-{a\over 4}(B+eA)_{\mu\nu}(B+eA)^{\mu\nu}
+{{\alpha}\over 2}
\epsilon_{\mu\nu\lambda}(B+eA)^\mu(B+eA)^{\nu\lambda}-{1\over{2g}}
B_\mu B^\mu,
\label{eqlb}
\end{equation}
where $\alpha =1/(4\pi)$ and $a=-1/(24\pi m)$.
In the present work, we treat the above $L_B$ in (\ref{eqlb}) as our
ML and the analysis actually begins here.
The ML has an invariance under local gauge transformation of $A_\mu$.
The
partion function corresponding to (\ref{eqlb}) in Lorentz gauge is
$$
Z=\int {\cal D}(A,B)\delta(\partial_\mu A^\mu)expi[
-{a\over 4}(B+eA)_{\mu\nu}(B+eA)^{\mu\nu}$$
\begin{equation}
+{{\alpha}\over 2}
\epsilon_{\mu\nu\lambda}(B+eA)^\mu(B+eA)^{\nu\lambda}-{1\over{2g}}
B_\mu B^\mu +J_\mu A^\mu+L_\mu B^\mu].
\label{eqz}
\end{equation}
The source current $J_\mu$ is conserved $(\partial_\mu J^\mu=0)$, but there is
no such restriction on the source current $L_\mu$. The classical
analysis including the spectrum of this model has already been discussed in
\cite{sg}. At present we are interested in the quantum Green Functions (GF)
$<A_\mu (x)A_\nu (y)>$ and $<B_\mu (x)B_\nu (y)>$
of the model. This will be achieved by the introduction of two {\it dual}
lagrangians, obtained by selective integration of the $A_\mu$ and $B_\mu$
fields respectively.
\vskip .5cm
{\bf III. Dual lagrangians}\\
\vskip .5cm

Our objective is to integrate out $A_\mu$ and $B_\mu$ separately from the ML.
Since $A_\mu$ is a gauge field, the computations have been carried out in a
particular (Lorentz) gauge, so that
$$\partial_\mu A^\mu=o.$$
The effective action is rewritten as,
$$
Z=\int{\cal D}(B)expi
[-{a\over 4}B_{\mu\nu}B^{\mu\nu}+{{\alpha}\over 2}
\epsilon_{\mu\nu\lambda}B^\mu B^{\nu\lambda}-{1\over{2g}}
B_\mu B^\mu +B_\mu L^\mu]$$
\begin{equation}
\int{\cal D}(A)expi[e^2A^\mu D_{\mu\nu}A^\nu
+eH_\mu A^\mu]
\label{eqza}
\end{equation}
where,
$$
D_{\mu\nu}=e^2[{{a\partial^2}\over 2}g_{\mu\nu}+(\zeta-{a\over 2})
\partial_\mu
\partial_\nu+(-\alpha)\epsilon_{\mu\nu\lambda}\partial^\lambda]
$$
$$
=e^2(P g_{\mu\nu}+Q\partial_\mu
\partial_\nu+R\epsilon_{\mu\nu\lambda}\partial^\lambda),
$$
$$H_{\mu}=
ea\partial^\nu B_{\nu\mu}+e\alpha\epsilon_{\mu\nu\lambda}
B^{\nu \lambda} +J_\mu. $$
$\zeta$ is the gauge parameter. Let us define,
\begin{equation}
D^{-1}_{\mu\nu}={1\over{e^2}}
[pg_{\mu\nu}+q\partial_\mu
\partial_\nu+r\epsilon_{\mu\nu\lambda}\partial^\lambda].
\label{eqd}
\end{equation}
Demanding $D^{\mu\nu}D^{-1}_{\nu\lambda}=g^\mu_\lambda$, we solve for $p$, $q$
and $r$ and get
$$
e^2p={P\over{P^2+R^2\partial^2}}={{2a\partial^2}\over{a^2(\partial^2)^2+
4\alpha^2\partial^2}},$$
$$
e^2r=-{R\over{P^2+R^2\partial^2}}={{4\alpha}\over{a^2(\partial^2)^2+
4\alpha^2\partial^2}},$$
$$
e^2q={{R^2-QP}\over{(P^2+R^2\partial^2)(P+Q\partial^2)}}
={{4\alpha^2-a(2\zeta -a)\partial^2}\over{\zeta\partial^2
(a^2(\partial^2)^2+
4\alpha^2\partial^2)}}.$$
After integration of $A_\mu$ we are left with,
$$Z_B(B_\mu)=\int{\cal D}(B_\mu)expi\int [(
-{a\over 4}B_{\mu\nu}B^{\mu\nu}+{{\alpha}\over 2}
\epsilon_{\mu\nu\lambda}B^\mu B^{\nu\lambda}-{1\over{2g}}
B_\mu B^\mu +B_\mu L^\mu )$$
\begin{equation}
+(-{i\over 4})(H^\mu D^{(-1)}_{\mu\nu}H^\nu)].
\label{eqzbb}
\end{equation}
After simplification we arrive at
$$
Z(B_\mu)=\int{\cal D}(B)exp(-{i\over 4})[{2\over g}B_\mu B^\mu+4B_\mu({1\over e}J^\mu-L^\mu)
+{{2a}\over {e^2}}J_\mu{1\over{a\partial^2+4\alpha^2}}J^\mu$$
\begin{equation}
-{{4\alpha}\over{e^2}} J_\mu{1\over{\partial^2
(a^2\partial^2+4\alpha^2)}}\epsilon_{\mu\nu\lambda}\partial^\nu J^\lambda]
\label{eqzb}
\end{equation}
Even though the source free expression of $Z(B_\mu )$ looks trivial with only
a contact term in $B_\mu$, the Green functions are quite involved, as we
will demonstrate in the next section.

In a similar way, we now carry out the $B_\mu$ integration. No gauge
fixing is required here. The starting partion function is reexpressed as
$$
Z_B(A_\mu)=\int{\cal D}(A)\delta(\partial_\mu A^\mu)
expi[-{{ae^2}\over 4}A^{\mu\nu}A_{\mu\nu}+{{\alpha e^2} \over 2}\epsilon
_{\mu\nu\lambda}A^\mu A^{\nu\lambda}+J_\mu A^\mu]$$
\begin{equation}
\int{\cal D}(B)expi
[-{a\over 4}B_{\mu\nu}B^{\mu\nu}+{{\alpha}\over 2}
\epsilon_{\mu\nu\lambda}B^\mu B^{\nu\lambda}-{1\over{2g}}
B_\mu B^\mu -{{ae}\over 2} B_{\mu\nu}A^{\mu\nu}+\alpha e\epsilon
_{\mu\nu\lambda}B^\mu A^{\nu\lambda} +L_\mu B^\mu].
\label{eqzbc}
\end{equation}
Once again we define the $B_\mu$-part as
$$\int{\cal D}(B)expi[B^\mu K_{\mu\nu}B^\nu
+h_\mu B^\mu]
$$
$$
K_{\mu\nu}=[({{a\partial^2}\over 2}-{1\over{2g}})g_{\mu\nu}+(-{a\over 2})
\partial_\mu
\partial_\nu+(-\alpha)\epsilon_{\mu\nu\lambda}\partial^\lambda]
$$
\begin{equation}
=P g_{\mu\nu}+Q\partial_\mu
\partial_\nu+R\epsilon_{\mu\nu\lambda}\partial^\lambda
\label{eqk}
\end{equation}

\begin{equation}
h_{\mu}=
ea\partial^\nu A_{\nu\mu}+e\alpha\epsilon_{\mu\nu\lambda}
A^{\nu \lambda} +L_\mu.
\label{eqh}
\end{equation}
Proceeding exactly as in the previous case, the required inverse operator is,
\begin{equation}
K^{-1}_{\mu\nu}=[pg_{\mu\nu}+q\partial_\mu
\partial_\nu+r\epsilon_{\mu\nu\lambda}\partial^\lambda].
\label{eqkpqr}
\end{equation}
$$
p={P\over{P^2+R^2\partial^2}}={{2g(ag\partial^2-1)}\over
{(ag\partial^2-1)^2+
4\alpha^2g^2\partial^2}},$$
$$
q={{R^2-QP)}\over{(P^2+R^2\partial^2)(P^2+Q\partial^2)}}=-{{2g^2
(4\alpha^2g+a^2g\partial^2)}\over{(ag\partial^2-1)^2+4\alpha^2g^2\partial^2}}
$$
$$
r=-{R\over{P^2+R^2\partial^2}}=
{{4\alpha g^2}\over
{(ag\partial^2-1)^2+
4\alpha^2g^2\partial^2}}.$$
The integration of $B_\mu$ leads to
$$Z_B(A_\mu)=\int{\cal D}(A_\mu)
\delta(\partial_\mu A^\mu)
expi[-{{ae^2}\over 4}A^{\mu\nu}A_{\mu\nu}+{{\alpha e^2} \over 2}\epsilon
_{\mu\nu\lambda}A^\mu A^{\nu\lambda}+J_\mu A^\mu]$$
\begin{equation}
+(-{i\over 4})(h^\mu K^{(-1)}_{\mu\nu}h^\nu)].
\label{eqzaa}
\end{equation}
The final result is the gauge invariant action,
$$
Z(A_\mu)=\int{\cal D}\delta (\partial_\mu A^\mu)exp(-i/4)[2e^2A_\mu
{{4\alpha^2g-a+a^2g\partial^2}\over{(ag\partial^2-1)^2+
4\alpha^2g^2\partial^2}}(\partial^2g^{\mu\nu}-\partial^\mu\partial^\nu)
A_\nu$$
$$+e^2A_\mu {{\{-4\alpha+8g(1-\alpha )
\partial^2(2\alpha^2g-a+a^2g\partial^2)\}}\over{(ag\partial^2-1)^2+
4\alpha^2g^2\partial^2}}\epsilon^{\mu\nu\lambda}\partial_\nu A_\lambda
$$
$$+4egA_\mu
{{4\alpha^2g-a+a^2g\partial^2}\over{(ag\partial^2-1)^2+
4\alpha^2g^2\partial^2}}(\partial^2g^{\mu\nu}-\partial^\mu\partial^\nu)
L_\nu$$
$$-8\alpha egL^\mu \epsilon_{\mu\nu\lambda}{1\over{(ag\partial^2-1)^2+
4\alpha^2g^2\partial^2}}\partial^\nu A^\lambda $$
$$-2g^2L^\mu\partial_\mu
(4\alpha^2g-a+a^2g\partial^2){1\over{(ag\partial^2-1)^2+
4\alpha^2g^2\partial^2}}\partial_\nu L^\nu $$
\begin{equation}
-4\alpha g^2L_\mu {1\over{(ag\partial^2-1)^2+
4\alpha^2g^2\partial^2}}\epsilon^{\mu\nu\lambda}\partial_\nu L_\lambda
+2gL_\mu(ag\partial^2-1){1\over{(ag\partial^2-1)^2+
4\alpha^2g^2\partial^2}}L^\mu -4J_\mu A^\mu].
\label{eqzda}
\end{equation}
The $\partial .A$ terms in (\ref{eqza}) have been displayed to show
the manifest gauge invariance but they are dropped later since Lorentz
gauge has been adopted.
Thus we have established an exact duality between the complicated
gauge invariant effective action in (\ref{eqzda}) and the more simpler
looking one in (\ref{eqzb}). It should be kept in mind that the duality
regarding the original fermion model in (\ref{eqlf}) and its bosonized
version in (\ref{eqz}),(as mentioned before), was not exact. On the
other hand, the just exposed duality is exact since they both originate
from the same master lagrangian in (\ref{eqz}) and no further
approximation are involved so far. The next task is to compute the
$A_\mu$ and $B_\mu$ GF's by exploiting the latter duality.
\vskip .5cm
{\bf IV. Green's functions}\\
\vskip .5cm
The sources present in (\ref{eqz}) allow us to formally define the GF's,
\begin{equation}
({1\over i})^2{{\delta^2 Z_B}\over{\delta J_\nu(y)\delta J_\mu (x)}}
\equiv <A^\mu(x)A^\nu(y)>,
\label{eqgf1}
\end{equation}
\begin{equation}
({1\over i})^2{{\delta^2 Z_B}\over{\delta L_\nu(y)\delta L_\mu (x)}}
\equiv <B^\mu(x)B^\nu(y)>.
\label{eqgf}
\end{equation}
But thanks to the duality we can replace $Z_B$ by $Z_B(A_\mu)$ or
$Z_B(B_\mu)$. Using $Z_B(B_\mu)$ in (\ref{eqzbc}) we find,
\begin{equation}
e^2<A^\mu(x)A^\nu(y)>
=<B^\mu(x)B^\nu(y)>+{i\over{a^2\partial^2+
4\alpha^2}}(ag^{\mu\nu}\delta(x-y)+{{2\alpha}\over{\partial^2}}\epsilon
^{\mu\nu\lambda}\partial_\lambda\delta(x-y)).
\label{eqg}
\end{equation}
In a similar way, performing the variation
$({1\over i})^2\delta^2 Z_B(A\mu)/(\delta L^\mu (x)
\delta L^\nu(y))$ on (\ref{eqgf}), we obtain the relation,
$$<B_\mu (x)B_\nu (y)>=[eg
{{(4\alpha^2g-a+a^2g\partial^2)}\over{(ag\partial^2-1)^2+
4\alpha^2g^2\partial^2}}\partial^2A_\mu(x)
$$
$$-2\alpha eg
{{1}\over{(ag\partial^2-1)^2+
4\alpha^2g^2\partial^2}}\epsilon_{\mu\alpha\beta}\partial^\alpha A^\beta(x)]
[eg
{{(4\alpha^2g-a+a^2g\partial^2)}\over{(ag\partial^2-1)^2+
4\alpha^2g^2\partial^2}}\partial^2A_\nu(y)
$$
$$-2\alpha eg
{{1}\over{(ag\partial^2-1)^2+
4\alpha^2g^2\partial^2}}\epsilon_{\nu\rho\sigma}\partial^\rho A^\sigma (y)]
$$
$$+i[-g^2{{(4\alpha^2g-a+a^2g\partial^2)}\over{(ag\partial^2-1)^2+
4\alpha^2g^2\partial^2}}\partial_\mu \partial_\nu \delta(x-y)
+2\alpha g^2{{1}\over{(ag\partial^2-1)^2+
4\alpha^2g^2\partial^2}}\epsilon_{\mu\nu\lambda}\partial^\lambda\delta(x-y)$$
\begin{equation}
+g(ag\partial^2-1)
{{1}\over{(ag\partial^2-1)^2+
4\alpha^2g^2\partial^2}}g_{\mu\nu}\delta(x-y)].
\label{eqgg}
\end{equation}
From the above
coupled Green functions, we can derive Green function identities containing
exclusively $A_\mu$ or $B_\mu$ fields, which are given below in a compact
form.
$$
e^2<A_\mu(x)A_\nu(y)>=e^2g^2M^{\mu\lambda}(x)M^{\nu\sigma}(y)<A_\lambda(x)
A_\sigma (y)>$$
$$+i[\{{a\over{a^2\partial^2+4\alpha^2}}+g(ag\partial^2-1)\phi\}
g^{\mu\nu}\delta(x-y)$$
\begin{equation}
-g^2(4\alpha^2g-a+a^2g\partial^2)\phi\partial^\mu\partial^\nu \delta(x-y)
+\{{{2\alpha}\over{\partial^2(a^2\partial^2+4\alpha^2)}}+2\alpha g^2\phi\}
\epsilon^{\mu\nu\lambda}\partial_\lambda\delta(x-y)],
\label{eqgfa}
\end{equation}
$$
<B_\mu(x)B_\nu(y)>=g^2M^{\mu\lambda}(x)M^{\nu\sigma}(y)<B_\lambda(x)
B_\sigma (y)>$$
$$+i[\{{{ag^2\partial^2}\over{a^2\partial^2
+4\alpha^2}}({{m_1^2}\over{\partial^2}}+m^2_3)
+g(ag\partial^2-1)\phi\}
g^{\mu\nu}\delta(x-y)$$
$$
-\{{{ag^2}\over{a^2\partial^2+4\alpha^2}}
({{m_1^2}\over{\partial^2}}+m^2_3)
+g^2(4\alpha^2g-a+a^2g\partial^2)\phi\}
\partial^\mu\partial^\nu \delta(x-y)$$
\begin{equation}
+\{{{2\alpha g^2}\over{(a^2\partial^2+4\alpha^2)}}
({{m_1^2}\over{\partial^2}}+m^2_3)
+2\alpha g^2\phi\}
\epsilon^{\mu\nu\lambda}\partial_\lambda\delta(x-y)].
\label{eqgfb}
\end{equation}
The operator $M^{\mu\lambda}(x)$ acts on $x$ and is defined below,
$$
M^{\mu\lambda}=m_1g^{\mu\lambda}+m_2\partial^\mu\partial^\lambda
+m_3\epsilon^{\mu\nu\lambda}\partial_\nu $$
$$
=(4\alpha^2g-a+a^2g\partial^2)\phi
[\partial^2g^{\mu\lambda}
-\partial^\mu\partial^\lambda ]
-2\alpha\phi\epsilon^{\mu\nu\lambda}\partial_\nu ,$$
\begin{equation}
\phi=[(ag\partial^2-1)^2+4\alpha^2g^2\partial^2]^{-1}.
\label{eqm}
\end{equation}
For the sake of comparison, we also write down the propagators for the Maxwell-
Chern-Simons model,
\begin{equation}
L_{MCS}={{-a}\over 4}A_{\mu\nu}A^{\mu\nu}+\alpha\epsilon^{\mu\nu\lambda}
A_\mu\partial_\nu A_\lambda +J_\mu A^\mu,~~~\partial.J=0,
\label{eqmcs}
\end{equation}
and the Maxwell-Chern-Simons-Proca model,
\begin{equation}
L_{MCSP}={{-a}\over 4}B_{\mu\nu}B^{\mu\nu}+\alpha\epsilon^{\mu\nu\lambda}
B_\mu\partial_\nu B_\lambda -{1\over{2g}}B_\mu B^\mu+L_\mu B^\mu .
\label{eqmcsp}
\end{equation}
The propagators are respectively,
\begin{equation}
<A^\mu(x)A^\nu(y)>_{MCS}=i({a\over{a^2\partial^2+4\alpha^2}}g^{\mu\nu}\delta(x-y)
+{{2\alpha}\over{\partial^2(a^2\partial^2+4\alpha^2)}}
\epsilon^{\mu\nu\lambda}\partial_\lambda\delta(x-y)),
\label{eqmcsa}
\end{equation}
$$
<B^\mu(x)B^\nu(y)>_{MCSP}={i\over{4\phi}}
[2g(ag\partial^2-1)g^{\mu\nu}\delta(x-y)-2g^2(4\alpha^2g+a^2g\partial^2-a)
\partial^\mu\partial^\nu\delta(x-y)
$$
\begin{equation}
+4\alpha g^2\epsilon^{\mu\nu\lambda}\partial_\lambda\delta(x-y)],
\label{eqmcspb}
\end{equation}
It is imperative to check the consistency of the above procedure
through a qualitative analysis of (\ref{eqgfa}) and (\ref{eqgfb}),
 by considering
limiting values of the parameters involved, specifically the fermion mass
$m$ and the Thirring coupling $g$.

The bosonization result was obtained in the large $m$ limit. In keeping
conformity with that, it is natural to keep only $O(m^{-1})$ terms and
hence $a^2\approx 0$. Indeed, this choice is by no means mandatory, since
the bosonized ML is an independent field theory by itself.
First we consider the large $g$ limit and keep terms of $O(a,g^{-1})$
only in the Green's functions.
$$
{{ae^2}\over{2\alpha^2g}}<A^\mu(x)A^\nu(y)>=e^2({a\over{2\alpha^2g}}-1)<
{{\partial^\mu\partial.}\over{\partial^2}}A(x)
[{{\partial^\nu\partial.}\over{\partial^2}}A(y)$$
$$
+A^\mu (x){{\partial^\nu\partial.}\over{\partial^2}}A(y)
+{{\partial^\mu\partial.}\over{\partial^2}}A(x)A^\nu(y)]>$$
$$
+{{e^2}\over{2\alpha g}}<[-A^\mu(x){{\epsilon^{\nu\rho\sigma}\partial_\rho}
\over{\partial^2}}A_\sigma (y)-
A^\nu(y){{\epsilon^{\mu\rho\sigma}\partial_\rho}
\over{\partial^2}}A_\sigma (x)+
{{\partial^\mu\partial .}\over{\partial^2}}A(x)
{{\epsilon^{\nu\rho\sigma}\partial_\rho}
\over{\partial^2}}b_\sigma (y)$$
$$
+{{\epsilon^{\mu\rho\sigma}\partial_\rho}\over{\partial^2}}A_\sigma(x)
{{\partial^\nu\partial .}
\over{\partial^2}}A(y)]>$$
\begin{equation}
+i[({a\over{4\alpha^2}}-{1\over{4\alpha^2g\partial^2}})g^{\mu\nu}\delta(x-y)
-({a\over{4\alpha^2}}+g){{\partial^\mu\partial^\nu}\over{\partial^2}}\delta
(x-y)+{1\over{\alpha\partial^2}}
\epsilon^{\mu\nu\lambda}\partial_\lambda
\delta(x-y).
\label{eqaag}
\end{equation}
$$
{a\over{2\alpha^2g}}<B^\mu(x)B^\nu(y)>=({a\over{2\alpha^2g}}-1)<
{{\partial^\mu\partial.}\over{\partial^2}}B(x)
$$
$$[{{\partial^\nu\partial.}\over{\partial^2}}B(y)
+B^\mu (x){{\partial^\nu\partial.}\over{\partial^2}}B(y)
+{{\partial^\mu\partial.}\over{\partial^2}}B(x)B^\nu(y)]>$$
$$
+{1\over{2\alpha g}}<[-B^\mu(x){{\epsilon^{\nu\rho\sigma}\partial_\rho}
\over{\partial^2}}B_\sigma (y)-
B^\nu(y){{\epsilon^{\mu\rho\sigma}\partial_\rho}
\over{\partial^2}}B_\sigma (x)+
{{\partial^\mu\partial .}\over{\partial^2}}B(x)
{{\epsilon^{\nu\rho\sigma}\partial_\rho}
\over{\partial^2}}B_\sigma (y)$$
$$
+{{\epsilon^{\mu\rho\sigma}\partial_\rho}\over{\partial^2}}B_\sigma(x)
{{\partial^\nu\partial .}
\over{\partial^2}}B(y)]>$$
\begin{equation}
+i[({a\over{2\alpha^2}}-{1\over{4\alpha^2g\partial^2}})g^{\mu\nu}\delta(x-y)
-({a\over{2\alpha^2}}+g){{\partial^\mu\partial^\nu}\over{\partial^2}}\delta
(x-y)+{1\over{2\alpha\partial^2}}(1+{a\over{2\alpha^2g}})
\epsilon^{\mu\nu\lambda}\partial_\lambda
\delta(x-y)].
\label{eqbbg}
\end{equation}
If we consider $g$-independent terms only, the $A_\mu$ and $B_\mu$
propagators become identical, but for an overall factor of ${1\over 2}$
in the $c$-number expressions. This is also the same as the $L_{MCS}$
propagator in (\ref{eqmcsa}). This is consistent with the fact that in
the master lagrangian in (\ref{eqz}), the difference between $A_\mu$
and $B_\mu$ disappears as $g\rightarrow\infty$ and the model tends to
$L_{MCS}$ in the composite field $(B_\mu+eA_\mu)$.

In case of small $g$, we have $\phi\approx 1$, $m_1\approx -2\alpha$
 and $m_3\approx -2\alpha$. This reproduces the following approximate
propagators,
\begin{equation}
<B^\mu (x)B^\nu (y)>\approx -ig^{\mu\nu}\delta(x-y),
\label{eqbap}
\end{equation}
\begin{equation}
e^2<A^\mu (x)A^\nu (y)>\approx i[{a\over{a^2\partial^2+4\alpha^2}}
g^{\mu\nu}\delta(x-y)+
{{2\alpha}\over{\partial^2(a^2\partial^2+4\alpha^2)}}
\epsilon^{\mu\nu\lambda}\partial_\lambda\delta(x-y).
\label{eqaap}
\end{equation}
Notice that the $A_\mu $-propagator in (\ref{eqbap}) is identical to
the MCS propagator in (\ref{eqmcsa}). This behaviour is justified in the limit of
small $g$ where the mass term for $B_\mu$ in (\ref{eqlb}) becomes infinite
and hence the $B_\mu$ field effectively decouples. This is also the
reason for trivial form of the $B_\mu$ propagator in (\ref{eqbap}).
\vskip .5cm
{\bf V: New dualities}
\vskip .5cm

We now consider two different ML's, which can be thought
of as large or small $e$-the electromagnetic coupling limit of the
original ML in (\ref{eqlb}) considered in the beginning.\\
(i). {\it Small $e$ limit:}\\
This procedure produces a different ML where
we drop $O(e^2)$ terms from starting
ML in  (\ref{eqlb}) and integrate $B_\mu$ first. The
resulting partition function is
$$
Z_1=\int{\cal D}(A)\delta(\partial .A)expi(J_\mu A^\mu)\int{\cal D}(B)
[-{{ae}\over 2}B_{\mu\nu}A^{\mu\nu} +e\alpha\epsilon_{\mu\nu\lambda}
B^\mu A^{\nu\lambda}+J_\mu A^\mu
$$
\begin{equation}
-{a\over 4}B_{\mu\nu}B^{\mu\nu}+{{\alpha}\over 2}
\epsilon_{\mu\nu\lambda}B^\mu B^{\nu\lambda}-{1\over{2g}}
B_\mu B^\mu +B_\mu L^\mu].
\label{eqz1}
\end{equation}
with $\partial .J=0$.
The result is
$$
Z_1(A_\mu )=\int{\cal D}(A)\delta(\partial .A)exp(-{i\over 4})[-4J_\mu A^\mu
+e^2A_\mu
\{2a^2g(\partial^2)^2(ag\partial^2-1)$$
$$
+4\alpha^2g\partial^2(ag\partial^2+1)\phi \}(\partial^2 g^{\mu\nu}
-\partial^\mu\partial^\nu)A_\nu $$
$$
+e^2
A_\mu
\{4g\partial^2(\alpha a^2g\partial^2-2a-4\alpha^2g)\}\phi
\epsilon^{\mu\nu\lambda}\partial_\nu A_\lambda
$$
$$+4eg(A_\mu
\{4\alpha^2g-a+a^2g\partial^2\}\phi (\partial^2 g^{\mu\nu}-
\partial^\mu\partial_\nu )A_\nu
$$
$$
-8\alpha egL^\mu \epsilon_{\mu\nu\lambda}{1\over{(ag\partial^2-1)^2+
4\alpha^2g^2\partial^2}}\partial^\nu A^\lambda $$
$$
-2g^2L^\mu\partial_\mu
(4\alpha^2g-a+a^2g\partial^2){1\over{(ag\partial^2-1)^2+
4\alpha^2g^2\partial^2}}\partial_\nu L^\nu $$
\begin{equation}
-4\alpha g^2L_\mu {1\over{(ag\partial^2-1)^2+
4\alpha^2g^2\partial^2}}\epsilon^{\mu\nu\lambda}\partial_\nu L_\lambda
+2gL_\mu(ag\partial^2-1){1\over{(ag\partial^2-1)^2+
4\alpha^2g^2\partial^2}}L^\mu ].
\label{eqa1a}
\end{equation}
In case of $A_\mu$ integration, the gauge fixing is really not necessary
since $A_\mu$ appears only linearly in the action. The $A_\mu$ equation
of motion produces the relation
\begin{equation}
-ea\partial^\nu B_{\nu\mu}+e\alpha\epsilon_{\mu\nu\lambda}B^{\nu\lambda}
+J_\mu =0.
\label{eqz1a}
\end{equation}
Substituting the above relation in the action, the result is\\
\begin{equation}
Z_1(B_\mu)=\int{\cal D}(B)expi[-{a\over 2}B_{\mu\nu}B^{\mu\nu}-{1\over{2g}}B_\mu
B^\mu-{1\over{2e}}B_\mu J^\mu+B_\mu L^m ].
\label{eqz1b}
\end{equation}
Without the sources, this is the Proca Model.
However, the Green's functions are,
$$
4e^2<A^\mu(x)A^\nu(y)>=e^2g^2M^{\mu\lambda}(x)M^{\nu\sigma}(y)<A_\lambda(x)
A_\sigma (y)>
$$
\begin{equation}
+i\{g(ag\partial^2)\phi g^{\mu\nu}\delta(x-y)
-g^2(4\alpha^2g-a+ag\partial^2)\phi\partial^\mu\partial^\nu\delta(x-y)
+2\alpha g^2\phi\epsilon^{\mu\nu\lambda}\partial_\lambda\delta(x-y)\},
\label{eqaa1}
\end{equation}
where the operator $M^{\mu\nu}(x)$ is the same as in (\ref{eqm}). The
$B_\mu$ Green function is obtained from the above one by replacing $A_\mu$
by $B_\mu/(2e)$. The reason is that the variation by $-i\delta/\delta
J_\mu$ reproduces $A_\mu$ and $-B_\mu/(2e)$ when operated upon
$Z_1$ and $Z_1(B_\mu )$ respectively.\\
(ii) {\it Large $e$ limit:}\\
We drop the $e$-independent quadratic $B$-terms from the original ML.
$$
Z_2=\int{\cal D}(A)\delta(\partial_\mu A^\mu)
expi[-{{ae^2}\over 4}A^{\mu\nu}A_{\mu\nu}+{{\alpha e^2} \over 2}\epsilon
_{\mu\nu\lambda}A^\mu A^{\nu\lambda}+J_\mu A^\mu]$$
\begin{equation}
\int{\cal D}(B)expi
[-{1\over{2g}}
B_\mu B^\mu -{{ae}\over 2} B_{\mu\nu}A^{\mu\nu}+\alpha e\epsilon
_{\mu\nu\lambda}B^\mu A^{\nu\lambda} +L_\mu B^\mu].
\label{eqz2}
\end{equation}
After $B_\mu$ integration, the result is
$$
Z_2(A_\mu )=\int{\cal D}(A)\delta(\partial_\mu A^\mu)exp(-i)[e^2A_\mu
({a\over 2}+{{a^2g}\over 2}\partial^2-2\alpha^2g) (\partial^2g^{\mu\nu}
-\partial^\mu\partial^\nu)A_\nu+2\alpha age^2A_\mu\partial^2\epsilon^{\mu\nu\lambda}
\partial_\nu A_\lambda $$
\begin{equation}
+eagL_\mu\partial_\nu A^{\nu\mu}+{g\over 2}L_\mu L^\mu
+2\alpha egL^\mu\epsilon^{\mu\nu\lambda}\partial_\nu A_\lambda ].
\label{eqz2b}
\end{equation}
Integrating out the $A_\mu$ field in the Lorentz gauge gives,
$$Z_2(B_\mu )=\int{\cal D}(B)exp(-i)[{a\over 2}(B_\mu\partial^2 B^\mu
-B_\mu\partial^\mu\partial^\nu B_\nu )+{1\over{2g}}B_\mu B^\mu
+B_\mu({1\over e}J^\mu-L^\mu)
+{a\over {2e^2}}J_\mu{1\over{a\partial^2+4\alpha^2}}J^\mu$$
\begin{equation}
-{{\alpha}\over{e^2}} J_\mu{1\over{\partial^2
(a^2\partial^2+4\alpha^2)}}\epsilon_{\mu\nu\lambda}\partial^\nu J^\lambda].
\label{z2a}
\end{equation}
Without the sources, this is the Chern-Simons-Proca model.
The Green's functions in this case are
$$
<A^\mu(x)A^\nu(y)>={1\over{e^2}}N^{\mu\beta}(x)N^{\nu\lambda}(y)
<A_\beta(x)A_\lambda(y)>$$
\begin{equation}
+{i\over{e^2}}\{(g+{a\over{a^2\partial^2+4\alpha^2}})
g^{\mu\nu}\delta(x-y)-{{2\alpha}\over{\partial^2(a^2\partial^2+4\alpha^2)}}
\epsilon^{\mu\nu\lambda}\partial_\lambda\delta(x-y)\},
\label{eqz3a}
\end{equation}
$$
<B^\mu(x)B^\nu(y)>={1\over{e^2}}N^{\mu\beta}(x)N^{\nu\lambda}(y)
<B_\beta(x)B_\lambda(y)>$$
\begin{equation}
+i\{gg^{\mu\nu}\delta(x-y)+{1\over{e^2}}
N^{\mu\beta}(x)N^{\nu\lambda}(y)
({a\over{a^2\partial^2+4\alpha^2}}g_{\beta\lambda}\delta(x-y)
+{{2\alpha}\over{\partial^2(a^2\partial^2+4\alpha^2)}}
\epsilon_{\beta\lambda\sigma}\partial^\sigma\delta(x-y))\},
\label{eqz3b}
\end{equation}
The operator $N^{\mu\nu}$ is defined as
$$N^{\mu\nu}=eag\partial^2g^{\mu\nu}-eag\partial^\mu\partial^\nu
-2\alpha eg\epsilon^{\mu\nu\lambda}\partial_\lambda.$$
\vskip .5cm
{\bf VI: Conclusions}\\
\vskip .5cm

Bosonized version of the gauged, massive Thirring model is analysed
exhaustively at the level of effective action. Duality invariance is
used to construct exactly equivalent gauge invariant and non-invariant
models. The correlation functions of the bosonized models are studied.
Validity of this new scheme to achieve this
is tested in different coupling regimes.
Depending on the relative strengths of the electromagnetic and Thirring
couplings, some new dual models, including their propagators, are also presented.\\

\vskip .5cm
{\bf Acknowledgements}\\
\vskip .5cm

It is a pleasure to thank Dr. Rabin Banerjee for a number of helpful
discussions. I am also grateful to Professor C. K. Majumdar, Director,
S. N. Bose National Centre for Basic Sciences,
Calcutta, for allowing me to use the
Institute facilities.

\end{document}